\documentclass[fleqn,12pt,twoside]{article}
\usepackage{espcrc1}

\usepackage{graphicx}
\usepackage[figuresright]{rotating}

\usepackage{epsfig}

\newcommand{\AmS}{{\protect\the\textfont2
  A\kern-.1667em\lower.5ex\hbox{M}\kern-.125emS}}

\title{  $NN(^1S_0)$ pairs in $^3$He  
 and in  $p^3$He  backward elastic scattering
}
\author{Yu.N. Uzikov\address[MCSD]{Joint Institute for Nuclear Research, 
        Dubna, Moscow region, Russia, 141980 }%
        \thanks {Permanent address: Kazakh National University,
        Institute for Experimental and Theoretical Physics,
        480078 Almaty, Kazakhstan; Supported by BMBF (grant KAZ-02/001 and
        Heisenberg-Landau program)} and
        J. Haidenbauer\address{ Institut f\"ur Kernphysik,
 Forschungszentrum J\"ulich, D-52425 J\"ulich, Germany}}

\begin{document}

\maketitle

\begin{abstract}
{It is shown that the shoulder observed in the cross section of
 $p^3He$ backward elastic scattering at proton beam energies of 0.4 -- 0.7 GeV
 can be explained by the mechanism of virtual $\pi$-meson production
 with a dominant contribution from the subprocesses  
 $pd^*\to ^3$He$\,\pi^0$ and $p(pp)\to^3$He$\,\pi^+$, where $d^*$ 
 denotes the singlet deuteron and the $pp$ pair is in the $^1S_0$ state. 
 At high energies, $1-3$ GeV, the mechanism of $np$-pair transfer dominates
and probes high-momentum $NN$($^1S_0$) correlations in $^3$He.
 }
\vspace{1cm}
\end{abstract}


 
\begin{figure}[t]
\mbox{\epsfig{figure=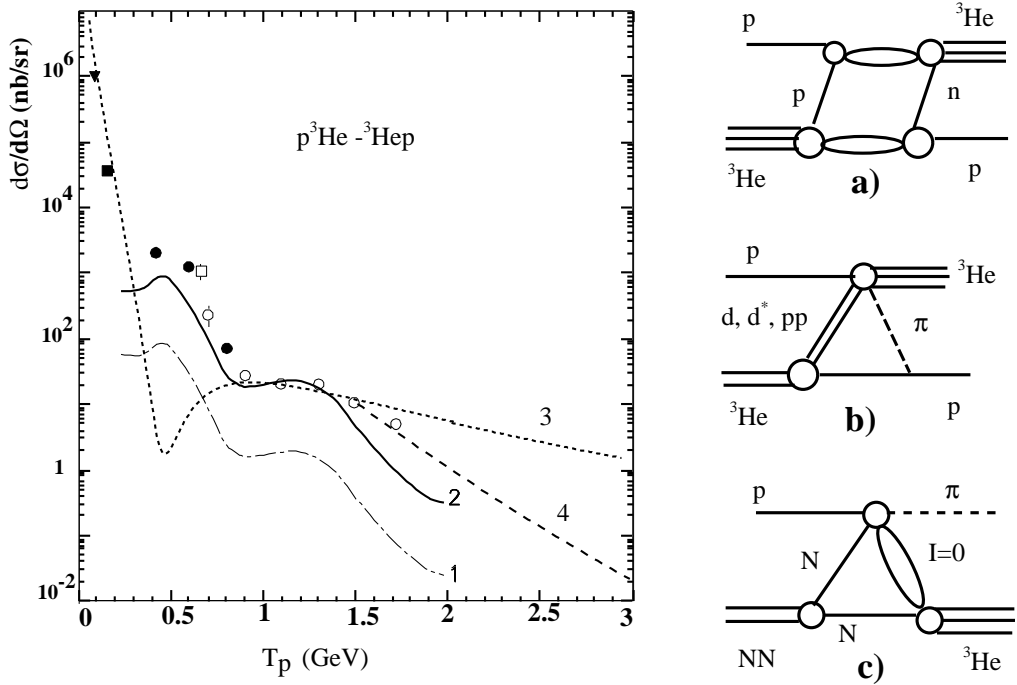,height=0.4\textheight, clip=}}
\caption{Right panel: The ST ({\it a}) and OPE ({\it b}) mechanisms 
of elastic $p^3He$ scattering, and the spectator mechanism of
p(NN)$\to^3$He $\pi$  ({\it c}).
Left panel: 
Cms cross section of p$^3$He$\to ^3$He\,p 
 at the scattering angle $\theta_{cm}=180^\circ$ versus 
 the proton beam energy. 
 Calculations on the basis of the OPE model with the $\pi NN$ cut-off 
 momentum $\Lambda_\pi= 1.3$ GeV/c: 1 -- for $d$ in the intermediate state,
 2 -- $d+d^* +pp$.
The result for the nondistorted ST cross section is given by the dotted
line 3 (divided by factor of 3 to normalize on the experiment at 1.5 GeV). 
The dashed curve 4 shows the slope for the counting rule $d\sigma/dt\sim
s^{-22}$ \protect\cite{brodsky}.
Experimental data are from Refs.
\protect\cite {berthet81} ($\circ$), \protect\cite{langevi} (filled square),
\protect\cite{komarov} (open square),
\protect\cite{frascaria}
 ($\bullet$), and \protect\cite{votta} (filled triangle).} 
\label{cross2}
\end{figure}

 Over the past few years $p^3$He backward elastic scattering  
 has been investiged \cite{blu96,uznpa98}
 on the basis of the DWBA using a $3N$ bound-state wave function 
 obtained from solving the Faddeev
 equations for the Reid RSC $NN$ potential. 
 Those studies suggested that this process at beam
 energies $T_p>1$~GeV can give unique information
 about the high momentum component of the $^3$He wave function
 $\varphi^{23} ({\bf q}_{23},{\bf p}_1)$, and specifically for
 high relative momenta, $q_{23}>0.6$ GeV/c, 
 of the nucleon pair $\{23\}$ in the $^1S_0$ state
 and low momenta of the nucleon "spectator"  $p_1<0.1$ GeV/c.
 Here $\varphi^{23}$ is the first Faddeev
 component of the full wave function of $^3$He,
 $\Psi(1,2,3)=\varphi^{23}+\varphi^{31}+\varphi^{12}$.
  The calculations presented in Refs. 
\cite{blu96,uznpa98}
 demonstrate the dominance of 
 the mechanism of sequential transfer (ST) of the
 proton-neutron ($np$) pair (Fig.\ref{cross2}a) 
 over a wide range of beam energies, 
 $T_p=0.1-2$ GeV, except for the region of the ST dip at around 0.3 GeV.  
 Other mechanisms of two-nucleon transfer,
 such as deuteron exchange, 
 non-sequential $np$ transfer \cite{blu96},
 and direct $pN$ scattering \cite{landau}
 involve very high internal momenta in the $^3{\rm He}$ wave function
in $q_{23}$ as well as in $p_1$ and, as a consequence, give much smaller 
 contributions. 
 As is shown here (see for details Ref. \cite{uzhaid}),
 the region of the ST dip (0.4-0.7 GeV) is dominated by the triangle 
diagrams of one pion exchange (OPE)
 with the subprocesses $pd^*\to\,^3{\rm He}\,\pi^0$ and
 $p(pp)\to\,^3{\rm He}\,\pi^+$ (Fig.~\ref{cross2}b),
 where $d^*$ and $pp$ are the spin-singlet $^1S_0$ deuteron and
 diproton, respectively. 


In Ref. \cite{uznpa98} the deuteron contribution to the cross section
 of $p^3$He scattering 
 within the OPE mechanism is expressed via the experimental
 cross section of the reaction $pd\to\,^3{\rm He}\,\pi^0$, without
 elaboration of its concrete mechanism. 
 In order to calculate the contribution of the meson production on the
 $d^*$ and on the diproton in $^3$He, we use
the $d^*+p$ and $(pp)+n$ configurations of $^3$He calculated in 
Refs. \cite{sciavilla,germwilkin} 
and adopt the spectator mechanism for the
subprocesses $p(NN)_{s,t}\to\,^3{\rm He}\,\pi$ (Fig.\ref{cross2}c)
on the singlet ($s$) and triplet ($t$) $NN$ pairs.
The analysis of the reaction $pd\to ^3He\pi^0$ 
 at $T_p<$1 GeV \cite{germwilkin,lagetlec}
 and the existing data on the inverse reactions of pion 
 capture, i.e. $\pi^{+}\,^{3}{\rm He}\to ppp$ and
 $\pi^{-}\,^{3}{\rm He}\to pnn$,
 confirm this mechanism, cf. the discussion in Ref. \cite{uzhaid}.

 Because of the specific spin structure of the OPE amplitude
 there is no interference between the triplet 
($M_d$) and singlet ($M_{d^*}+M_{pp}$)
 amplitudes in the spin-averaged sum: 
 \begin{equation}
 \label{averagedm}   
{\overline {|M_d+M_{d^*}+M_{pp}|^2}}=
|K|^2 \left \{ {\overline {|G_d\,T_d}|^2} 
 +\frac{1}{3}
 {\overline {|G_{d^*}(T_{d^*}+
2\,T_{pp})|^2}}  
 \right \}.
\end{equation}
 Here $T_\alpha$ ($\alpha=d$,$d^*$ and $pp$) is the meson production
 amplitude p(NN)$_\alpha \to ^3$He$\pi$ 
 and $G_\alpha$ is the structure factor \cite{uzhaid}.
 The factors $\frac{1}{3}$ and $2$ in the second term in the curly brackets 
 of Eq.~(\ref{averagedm}) are combinations of isospin coefficients. 
 Furthermore, 
 we assume that the subprocess $pN\to (NN)_t\,\pi$ dominates in 
 the upper vertex of the diagram in Fig.~\ref{cross2}c and that the  
 amplitude $pN\to (NN)_s\,\pi$ is negligible. This is true in 
 the $\Delta$-region \cite{uzwilkin}. 
 With this approximation
 there is a relation between the amplitudes of the processes
 $pd^*\to\,^3{\rm He}\,\pi^0$ and 
 $p(pp)_s\to\,^3{\rm He}\,\pi^+$ which follows from isospin invariance,
 namely $ T_{pp}=2\,T_{d^*}$.
 After that one gets 
the relation
$ {\overline {|T_{d^*}+2T_{pp}|^2}}=25 {\overline {|T_d|^2}}.$
 Therefore the combined contribution of $T_{d^*} + T_{pp}$
 is significantly larger than that
 of the deuteron and can be expressed via the experimental cross
 section of the reaction $pd\to\,^3{\rm He}\,\pi^0$
 taken here from Ref. \cite{berthet85}.
  Distortions are included in the factor $K$.
  

 The results of our calculation are shown in Fig.~\ref{cross2}.
 Evidently, the OPE model with deuteron exchange (curve 1)
 yields a reasonable description of the energy dependence
 of the cross section for $T_p= 0.4 - 1.5$~GeV. However, due to 
 distortions, taken into account here in eikonal approximation, 
 the $d$ contribution turns out to be  
 one order of magnitude smaller than the data.
 The contributions of the singlet deuteron $d^*$ and of the $pp$ pair
 bring the calculation in qualitative agreement with the data (curve 2).  

 One can see from curve 3 in Fig.~\ref{cross2} 
 that the ST mechanism, calculated
 here with the $3N$ wave function \cite{vbaru} resulting from the 
 CD Bonn potential, is significant at beam energies $T_p=0.9-1.5$~GeV and 
 it definitely dominates at low ($T_p<0.3$~GeV) and high ($T_p>1.5$~GeV)
 energies.
 Since at sufficiently high relative $NN$ momenta the concept of 
 having individual nucleons inside nuclei is expected to fail,
 we show in Fig.\ref{cross2}, for comparison, also
 the slope of the cross section as it follows from quark counting
 rules \cite{brodsky} for direct (without baryon exchanges) mechanism.
 Obviously, this slope differs from the one of the ST and OPE mechanisms.  
%

 In conclusion, our calculations suggest that the shoulder in 
 the $p^3{\rm He}\to\,^3{\rm He}p$ cross section 
 at 0.4--0.6 GeV is mainly due to the OPE mechanism with the singlet
 $NN(^1S_0)$ pairs in $^3$He.
 A measurement of spin observables, planned 
 at 0.2-0.4 GeV \cite{Hatan} could give additional information here.

\vfill 

\end{document}